\documentclass{article}
\usepackage{graphicx}
\usepackage{amsmath,graphicx,amssymb,amsthm,bm}
\usepackage{color}
\usepackage{float}
\usepackage{multicol}
\usepackage{makecell}
\usepackage{subcaption}
\usepackage{amsfonts}
\usepackage{mathtools}
\usepackage{multirow}
\usepackage{booktabs}
\usepackage{array}
\usepackage{bm}
\usepackage{nicefrac}
\usepackage{tabularx}
\usepackage{booktabs}
\usepackage{xcolor}
\usepackage{extarrows}
\usepackage{balance}
\usepackage{spconfa4}

\newcommand{\eref}[1]{(\ref{#1})}
\newcommand{\sref}[1]{Section~\ref{#1}}
\newcommand{\fref}[1]{Fig.~\ref{#1}}

\copyrightnotice{978-1-6654-6867-1/22/\$31.00~\copyright2022 IEEE}
                   
\title{Meta-Learning for Adaptive Filters with\\Higher-Order Frequency Dependencies}
\name{Junkai Wu$^{\sharp}$\thanks{Partially funded by NIFA award 2020-67021-32799 $^1$https://jmcasebeer.github.io/metaaf/higher-order} \qquad Jonah Casebeer$^{\sharp}$ \qquad 
Nicholas J. Bryan$^{\flat}$ \qquad 
Paris Smaragdis$^{\sharp}$ 
}%
\address{
$^\sharp$ University of Illinois at Urbana-Champaign, $^\flat$ Adobe Research\\
}
\begin{document}
%
\maketitle
\begin{abstract}
Adaptive filters are applicable to many signal processing tasks including acoustic echo cancellation, beamforming, and more.
Adaptive filters are typically controlled using algorithms such as least-mean squares~(LMS), recursive least squares~(RLS), or Kalman filter updates. Such models are often applied in the frequency domain, assume frequency independent processing, and do not exploit higher-order frequency dependencies, for simplicity.
Recent work on meta-adaptive filters, however, has shown that we can control filter adaptation using neural networks without manual derivation, motivating new work to exploit such information.
In this work, we present higher-order meta-adaptive filters, a key improvement to meta-adaptive filters that incorporates higher-order frequency dependencies.
We demonstrate our approach on acoustic echo cancellation and develop a family of filters that yield multi-dB improvements over competitive baselines, and are at least an order-of-magnitude less complex. Moreover, we show our improvements hold with or without a downstream speech enhancer.
\end{abstract}
\section{Introduction}
Adaptive filters~(AFs) are broadly useful for numerous audio tasks such as acoustic echo cancellation, equalization, and multi-channel denoising or beamforming. AFs are typically defined as linear filters with time-varying filter weights that are computed by solving an online optimization problem via additive update rules. Example hand-derived AF algorithms include least-mean squares~(LMS)~\cite{widrow1960adaptive}, normalized LMS~(NLMS), recursive least squares~(RLS), and Kalman filters~(KF)~\cite{Widrow1985, mathews1991adaptive, haykin2008adaptive, apolinario2009qrd, rabiner2016theory}. Early AFs used time-domain filters~\cite{widrow1960adaptive}, but were quickly replaced with (multi-) block-frequency domain filters~\cite{mansour1982unconstrained, soo1990multidelay, haykin2008adaptive}, which often assume frequency-independent processing for simplicity. 

When we survey further improvements to AFs with a focus on acoustic echo cancellation~(AEC)~\cite{benesty2001advances, hansler2005acoustic}, numerous improvements have been proposed. For example, near-end signal models have been proposed for handling simultaneous far-end and near-end activity (double-talk)~\cite{gay1998efficient, valin2007adjusting, haubner2021noise} as well as state-space formulations~\cite{enzner2006frequency, malik2010online, kuech2014state, yang2017frequency}. A few works~\cite{avargel2007system, valero2015state} also propose to use higher-order frequency dependencies for AEC AFs, likely motivated by the success of multivariate statistics for source separation~\cite{buchner2004trinicon, kim2006blind}, but showed varied performance improvement with increased complexity.

\begin{figure}[!t]
    \centering
    \includegraphics[trim=0 0 0 .75cm, clip, width=.9\linewidth]{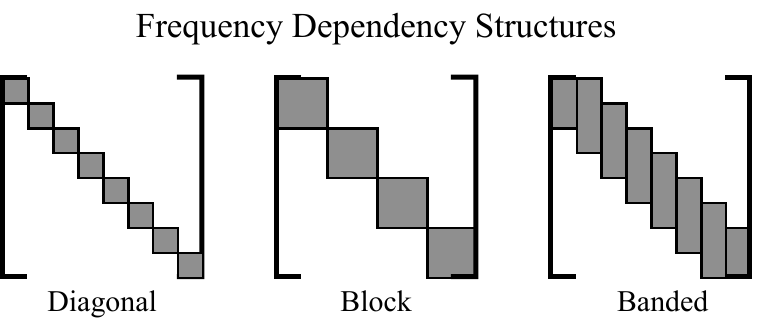}
    \vspace{-5mm}
    \caption{Frequency dependency structures for meta-adaptive filters. (Left) Diagonal. (Center) Block. (Right) Banded.}
    \label{fig:ho-metaaf}
    \vspace{-3mm}
\end{figure}

More recent improvements to AFs include data-driven methods. Such methods include model-based methods~\cite{casebeer2021nice, haubner2021end, haubner2022deep}, which use deep neural networks~(DNN) to estimate signal statistics for existing signal models and model-free approaches~\cite{casebeer2021auto, casebeer2022metaaf}. Of particular interest is Meta-AF~\cite{casebeer2022metaaf}, an approach of using meta-learning to \emph{learn} adaptive filter update rules from data using neural networks. This approach was found to outperform several past methods for AEC, but still only learns frequency-independent update rules. It thus neglects higher-order frequency dependencies and requires one forward-pass of a neural network per frequency bin, making it more computationally complex than desired.

In this work, we present meta-adaptive filters with higher-order frequency dependencies --- a key extension to Meta-AFs that incorporates higher-order frequency dependencies into learned update rules as shown in~\fref{fig:ho-metaaf}. We demonstrate our approach on the task of AEC and develop a family of higher-order adaptive filters. Compared to Meta-AF~\cite{casebeer2021auto, casebeer2022metaaf}, we improve AEC performance by multiple decibels using an order of magnitude less floating point operations (FLOPs). We also compare against conventional AEC approaches and find our higher-order Meta-AF significantly outperforms all tested alternatives. Beyond this, we verify our proposed improvements hold with or without a down-stream DNN-based speech enhancer~(SE), showing the practical value of our approach. For reproducibility, we release our outputs, trained model weights, and code$^1$.
\def\time{{\mathrm{t}}}
\def\freq{\mathrm{k}}
\def\mic{{\mathrm{m}}}
\def\buffer{\mathrm{b}}
\def\frame{\tau}

\def\F{\mathbf{F}}
\def\I{\mathbf{I}}
\def\0{\mathbf{0}}
\def\1{\mathbf{1}}

\def\u{\mathbf{u}}
\def\U{\mathbf{U}}
\def\d{\mathbf{d}}
\def\D{\mathbf{D}}
\def\y{\mathbf{y}}
\def\e{\mathbf{e}}
\def\s{\mathbf{s}}
\def\w{\mathbf{w}}

\def\H{\mathsf{H}}
\def\T{\top}
\def\diag{\operatorname{diag}}

\def\L{\mathcal{L}}
\def\grad{\bm{\nabla}}

\def\bphi{\bm{\phi}}
\def\btheta{\bm{\theta}}

\section{Background}

\subsection{Adaptive filters}
\label{sec:background}

We define an AF as a time-varying linear filtering procedure $h_{\btheta[\tau]}$, with parameters $\btheta[\frame]$ computed by solving
\begin{equation}
    \hat{\btheta}[\tau] = \arg\min_{\btheta[\tau]} \L(h_{\btheta[\tau]},\cdots) \label{eq:general_af}
\end{equation}
via an additive update rule
\begin{equation}
    \btheta[\tau + 1] = \btheta[\tau] + \bm{\Delta}[\tau],
\label{eq:general_update}
\end{equation}
where signals are indexed by $\frame$, the AF loss $\L(\cdots)$ is a function of one or more signals, $\bm{\Delta}[\tau]$ is the AF update, and $\btheta[\tau]$ are filter parameters. 

We use frequency-domain AFs with overlap-save~(OLS) filtering denoted by $h_{\btheta[\frame]}(\cdot)$ with frequency coefficients $\btheta[\frame] = \w[\frame]\in \mathbb{C}^{K}$. We represent time-domain signals via an underline and frequency-domain without. In matrix notation, the OLS time-domain output $\underline{\y}[\frame]$ for an $R$ sample hop is computed via $\underline{\y}[\frame] = \mathbf{Z}_y \y[\frame] \in \mathbb{R}^R$, where $\y[\frame] = \diag(\u[\frame]) \mathbf{Z}_w \w[\frame] \in\mathbb{C}^{K}$, $\u[\frame]$ is the input signal, $\mathbf{Z}_w = \F_K \mathbf{T}_{R}^\T \mathbf{T}_R \F_K^{-1}\in\mathbb{C}^{K \times K}$, $\mathbf{Z}_y = \bar{\mathbf{T}}_{R} \F_K^{-1} \in \mathbb{C}^{R \times K}$ are anti-aliasing matrices, $\F_K$ is the $K$-point discrete Fourier transform matrix, $\mathbf{T}_{R} = [\I_{K-R}, \0_{K-R \times R}] \in \mathbb{R}^{K-R \times K}$ trims the last $R$ samples from a vector and $\bar{\mathbf{T}}_{R} = [\0_{K-R \times R}, \I_{K-R}] \in \mathbb{R}^{K-R \times K}$ trims the first $R$ samples. We set $R=K/2$. Succinctly, $\underline{\y}[\frame] = h_{\btheta[\frame]}(\u[\frame])$ and $\underline{\e}[\frame] = \underline{\d}[\frame] - \underline{\y}[\frame]$ where $\d$ is the desired response, and $\e$ is the error.

\subsection{Meta-adaptive filters}
Recent work has shown that AF update rules can be \emph{learned} from data using neural networks~\cite{casebeer2022metaaf}. To do so, a meta-learning formulation is used instead of~\eref{eq:general_af}, resulting in
\begin{equation}
    \hat{\bphi} = \arg\min_{\bphi} E_\mathcal{D}[\;  \L_M(\; g_{\bphi},  \L(h_{\btheta}, \cdots) \; ) \; ], 
\label{eq:meta}
\end{equation}
where $\L_M(\; g_{\bphi}, \L(h_{\btheta}, \cdots) \;)$ is the meta-loss that is a function of the AF loss, filter or optimizee $h_{\btheta[\frame]}(\cdot)$, optimizer neural network $g_{\bphi}(\cdot)$, and $E_\mathcal{D}$ represents expectation over dataset $\mathcal{D}$. Generally, this results in the update $\btheta[\frame + 1] = \btheta[\frame] + g_{\bphi}(\cdot)$,
where $g_{\bphi}(\cdot)$ is recurrent network parameterized by $\bphi$, applied per frequency $\freq$, and with configuration-dependent inputs. For a stateful optimizer, the update rules are
\begin{eqnarray}
(\bm{\Delta}_{\freq}[\frame], \bm{\psi}_{\freq}[\frame+1]) &=& g_{\bphi}(\bm{\xi}_{\freq}[\frame], \bm{\psi}_{\freq}[\frame]) \label{eq:full_neural_update1}\\
\btheta_{\freq}[\frame + 1] &=& \btheta_{\freq}[\frame] +  \bm{\Delta}_{\freq}[\frame],\label{eq:full_neural_update2}
\end{eqnarray}
where we index across $K$ frequencies using subscript $\freq$. The input is $\bm{\xi}_{\freq}[\frame] = [\nabla_{\freq}[\frame], \u_{\freq}[\frame], \d_{\freq}[\frame], \e_{\freq}[\frame], \y_{\freq}[\frame]]$ and the internal state $\bm{\psi}_{\freq}[\frame]$. Here, $\u_{\freq}[\frame]$ is the filter input, $\y_{\freq}[\frame]$ is the filter output, $\d_{\freq}[\frame]$ is the desired response, $\e_{\freq}[\frame]$ is the error, and $\nabla_{\freq}[\frame]$ are autodiff gradients of the AF loss w.r.t. $\btheta_\freq[\frame]$. The outputs are AF update $\bm{\Delta}_{\freq}[\frame]$ and a new internal state.

To learn network parameters $\bphi$, we use backpropagation-through-time and a meta-optimizer~(e.g. Adam) over $L$ steps to update $\bphi$ until convergence. For our meta-loss, we use the frame-accumulated meta-loss~\cite{casebeer2022metaaf}
\begin{eqnarray}
    \L_M(\cdots) &=& \ln E[\|\underline{\bar{\d}}[\frame] - \underline{\bar{\y}}[\frame]\|^2],
     \label{eq:metaloss}
\end{eqnarray}
which was found to be superior to alternatives, where $\underline{\bar{\d}}[\tau] = \mathrm{cat}(\underline\d[\tau], \cdots, \underline\d[\tau+L-1]) \in \mathbb{R}^{RL}$, $\underline{\bar{\y}}[\tau] = \mathrm{cat}(\underline\y[\tau], \cdots, \underline\y[\tau+L-1]) \in \mathbb{R}^{RL}$, and $\mathrm{cat}$ is the concatenation operator. For a full review and example code, please see~\cite{casebeer2022metaaf}.
\section{Higher-order meta-adaptive filters}

\subsection{Overview}
We build on Meta-AF~\cite{casebeer2022metaaf} and present a simple, but powerful extension to incorporate higher-order frequency dependencies when estimating filter updates. To learn frequency dependencies, we introduce learnable downsampling $\mathcal{S}$ and upsampling $\mathcal{U}$ layers before and after our optimizer network. The downsampling layer projects the per-frequency inputs $\bm{\xi}_{\freq}[\frame]$ into $C$ groups, where $C \leq K$. We run the optimizer $g_{\bphi}$ independently per coupled group $c$ instead of per frequency $k$ and use the upsampling layer to expand the group update $\bm{\Delta}_{\mathrm{c}}[\frame]$ to a per-frequency update. Formally, we modify~\eref{eq:full_neural_update1} and~\eref{eq:full_neural_update2} to be 
\begin{eqnarray}
(\bm{\Delta}_{\mathrm{c}}[\frame], \bm{\psi}_{\mathrm{c}}[\frame+1]) &=& g_{\bphi}(\; \mathcal{S}(\bm{\xi}[\frame])_{\mathrm{c}} \;,\; \bm{\psi}_{\mathrm{c}}[\frame] \;) \label{eq:ho_full_neural_update1}\\
\btheta_{\freq}[\frame + 1] &=& \btheta_{\freq}[\frame] +  \mathcal{U}(\bm{\Delta}_{\mathrm{c}})_{\freq}, \label{eq:ho_full_neural_update2}
\end{eqnarray}
where the modified network state $\bm{\psi}_{\mathrm{c}}[\frame+1]$ stores state per group. By applying $g_{\bphi}$ per group of frequencies, we can model interactions within a group and share state/computation within groups, significantly reducing computational cost. Each shaded square/rectangle in~\fref{fig:ho-metaaf} represents a group.  

\subsection{Dependency structures}
\label{sec:groups}
Our approach allows for arbitrary frequency dependencies by imposing structure into the up-/down- sampling layers. We focus on three different forms of structure as shown in~\fref{fig:ho-metaaf}, including diagonal~(left), block~(middle), and banded~(right). On an intuitive level, each coupling structures implies a different inter-frequency covariance matrix. Diagonal corresponds to frequency independent processing, while block/banded model higher order relationships and allow information sharing across frequency groups. Larger groups model more interactions, but at the cost of sharing a single $\mathrm{H}$ dimensional state. Thus, there is a trade-off between group size, state size, performance, and efficiency. We describe three potential dependency structures below. 

\textit{Diagonal:}
For diagonal frequency dependencies, we set $\mathcal{S}$ and $\mathcal{U}$ to be dense layers operating identically on each frequency. We use this configuration as a baseline as it defaults to~\cite{casebeer2022metaaf}. The complexity of $g_{\bphi}$ is $\mathcal{O}(\mathrm{H}^2)$, with $K$ executions per frame resulting in a total complexity of $\mathcal{O}(K\mathrm{H}^2)$.

\textit{Block:}
For block frequency dependencies, we reshape each of the $K$ per-frequency network inputs $\bm{\xi}_{\freq}[\frame] = [\nabla_{\freq}[\frame], \u_{\freq}[\frame], \d_{\freq}[\frame], \e_{\freq}[\frame], \y_{\freq}[\frame]] \in \mathbb{C}^{5}$ into $C$ per-group features $\bm{\xi}_{c}[\frame] \in \mathbb{C}^{5\mathrm{B}}$ or a $\mathrm{C} \times 5\mathrm{B}$ matrix, where $C = K/B$ and $B$ is the group size.

We then apply a dense layer on the latter dimension to produce a $\mathrm{C} \times \mathrm{H}$ output and apply the optimizer separately to each of the $\mathrm{C}$ columns. Thus, we impose a non-overlapping block-group structure and enable information and computation sharing within groups. This is reminiscent of sub-band processing~\cite{kuech2014state}. The cost of $g_{\bphi}$ is $\mathcal{O}(\mathrm{H}^2)$, the up/down sampling layers cost $\mathcal{O}(\mathrm{B}\mathrm{H})$, and the number of executions per frame is $\frac{K}{\mathrm{B}}$, for a total of $\mathcal{O}(\frac{K}{\mathrm{B}}(\mathrm{H}^2 + \mathrm{B}\mathrm{H}))$.

\textit{Banded:}
For banded frequency dependencies, we modify the block-reshape operation described above to return overlapping groups of $\mathrm{B}$ frequencies, and retain all other block dependency operations. By doing so, we enable information and computation sharing across overlapping groups of frequencies and better model adjacent frequency relationships. By increasing and decreasing the overlap, we modulate the number of adjacent frequencies. In this work, we set the overlap to $\frac{\mathrm{B}}{2}$. This style of dependencies was explored in past work~\cite{avargel2007system, valero2015state}. The complexity of $g_{\bphi}$ is $\mathcal{O}(\mathrm{H}^2)$, the up/down sampling layers cost $\mathcal{O}(\mathrm{B}\mathrm{H})$, and the number of executions per frame is $2\frac{K}{\mathrm{B}}$ for a total of $\mathcal{O}(\frac{K}{\mathrm{B}}(\mathrm{H}^2 + \mathrm{B}\mathrm{H}))$.

Practically, we implement all dependency structures using standard deep learning operations. We implement the downsampling layer with a 1-D convolution and the upsampling layer with a transposed convolution. We configure different strategies with different filter sizes~($\mathrm{B}$), and stride sizes. A filter size of $\mathrm{B}=1$ with stride one implements diagonal, a filter size where $\mathrm{B} > 1$ with stride of $\mathrm{B}$ implements block, and a filter size of $\mathrm{B} > 1$ and stride of $\mathrm{B}/2$ implements banded.
\section{Experimental Design} 
\begin{figure}[!t]
    \centering
    \includegraphics[trim=.5cm .5cm 1cm .2cm, clip, width=.99\linewidth]{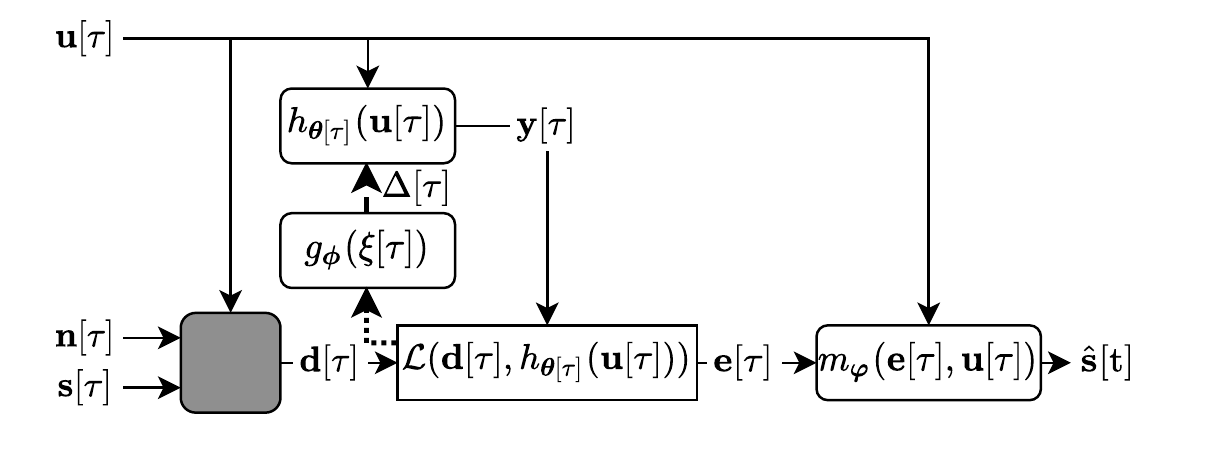}
    \vspace{-3mm}
    \caption{Higher-order Meta-AF for AEC with a DNN speech enhancer. The shaded box represents an unknown system.}
    \label{fig:aec}
    \vspace{-2mm}
\end{figure}

We evaluate our method on AEC with double-talk, near/far -end noise, and nonlinearities. We implement an AEC~\fref{fig:aec} with an optional DNN speech enhancer~(DNN-SE), $m_{\bm{\varphi}}(\cdot)$ and benchmark  via objective, perceptual, and speed metrics.

\subsection{Acoustic echo cancellation problem formulation} 
To perform AEC, we fit a linear frequency-domain finite-impulse response filter to mimic an unknown echo path, output $\e[\frame]$ via the OLS filter $h_{\btheta[\frame]}(\cdot)$~(\sref{sec:background}), and use an AF loss of $\L(\cdot)=E[\|\e[\frame]\|^2]$. The signal-model is  $\underline{\d}[\time]=\sigma(\underline{\u}[\time])\ast\underline{\w}+\underline{\mathbf{n}}[\time]+\underline{\mathbf{s}}[\time]$ where $\underline{\mathbf{n}}$ is noise, $\underline{\mathbf{s}}$ is speech, and $\sigma(\cdot)$ is a loudspeaker nonlinearity.

\subsection{Configurations and baselines}
We compare block-frequency NLMS, RLS, KF~\cite{enzner2006frequency}, and diagonalized Meta-AF~\cite{casebeer2022metaaf} to higher-order Meta-AF with block and banded frequency dependency structures.

\subsection{Model details}
For all our higher-order dependency configurations, we set $g_{\bphi}$ to be a stack of two complex-valued gated recurrent units~(GRU) with hidden size $H$. For all dependency strategies, the output size of $\mathcal{S}$ is $H$, and the input size of $\mathcal{U}$ is $H$. We perform magnitude log-scaling to the inputs via $\ln(1 + |\bm{\xi}|)e^{j\angle\bm{\xi}}$ as in~\cite{casebeer2022metaaf}. For our enhancer $m_{\bm{\varphi}}(\e[\frame], \u[\frame])$, we follow~\cite{cutler2022AEC} and use a $2$-layer GRU with log-magnitude short-time Fourier transforms of the far-end, $\u[\frame]$ and the AEC output $\e[\frame]$ as inputs with output $\hat{\s}[\frame] = \e[\frame] \odot \mathbf{M}[\frame]$ using magnitude mask, $\mathbf{M}[\frame]\in\mathbb{R}^K$ bounded with a Sigmoid function. $\odot$ is the hadamard product. $m_{\bm{\varphi}}(\cdot)$ is trained to remove noise and residual echo. We use JAX~\cite{bradbury2020jax}, Haiku~\cite{haiku2020github}, and the Meta-AF python package~\cite{casebeer2022metaaf}.

\subsection{Datasets}
We use the synthetic portion of the Microsoft AEC Challenge~\cite{cutler2022AEC}. It contains $10000$ pairs of $10\,$second scenes at $16\,$KHz, so we use $9000$, $500$, and $500$ for training, validation, and test. Each scene has double-talk, and optional near-end noise and loud-speaker nonlinearities. Double-talk occurs in the middle of every scene, so we apply a random circular shift. 

\subsection{Evaluation metrics}
We evaluate AEC performance using segmental echo return loss enhancement~(SERLE)~\cite{enzner2014acoustic}, and short-time objective intelligibility~(STOI)~\cite{taal2011algorithm}. When evaluating after the enhancer, we use STOI, and scale-invariant signal-to-distortion ratio~(SI-SDR)~\cite{le2019sdr}. With $\underline{\d}_{\u}[\frame]=\sigma(\underline{\u}[\time])\ast\underline{\w}$, SERLE is
\begin{equation}
    \sum_{\frame} \frac{10}{\mathrm{N}}\log_{10} \left(\; \| \underline{\d}_{\u}[\frame] \|^2 \;/\;(\| \underline{\d}_{\u}[\frame] - \underline{\y}[\frame] \|^2)\; \right),
\end{equation}
$\mathrm{N}$ is the number of frames, and we discard silent frames. 
$\operatorname{SI-SDR}$ uses $\mathbf{a} = (\underline{\hat{\mathbf{s}}}^\T \underline{\mathbf{s}})/\|\underline{\s}\|$ and is $\operatorname{SI-SDR}(\underline{\s}, \underline{\hat{\s}}) = 10 \cdot \log_{10}(\|\mathbf{a}\underline{\s}\|^2/\|\mathbf{a}\underline{\s} -\underline{\hat{\s}}\|^2)$.
To empirically evaluate speed, we use FLOPS and real-time-factor~(RTF)~(computation/time) to compliment our Big-O analysis. While there is debate on the utility of FLOPs for deep learning, we believe the measure is relevant for real-time low-power devices. RTF is also useful, but highly implementation and environment dependent.

\begin{figure*}[!t]
	\centering
	\includegraphics[trim=.2cm .2cm .25cm .25cm, clip, width=.97\linewidth]{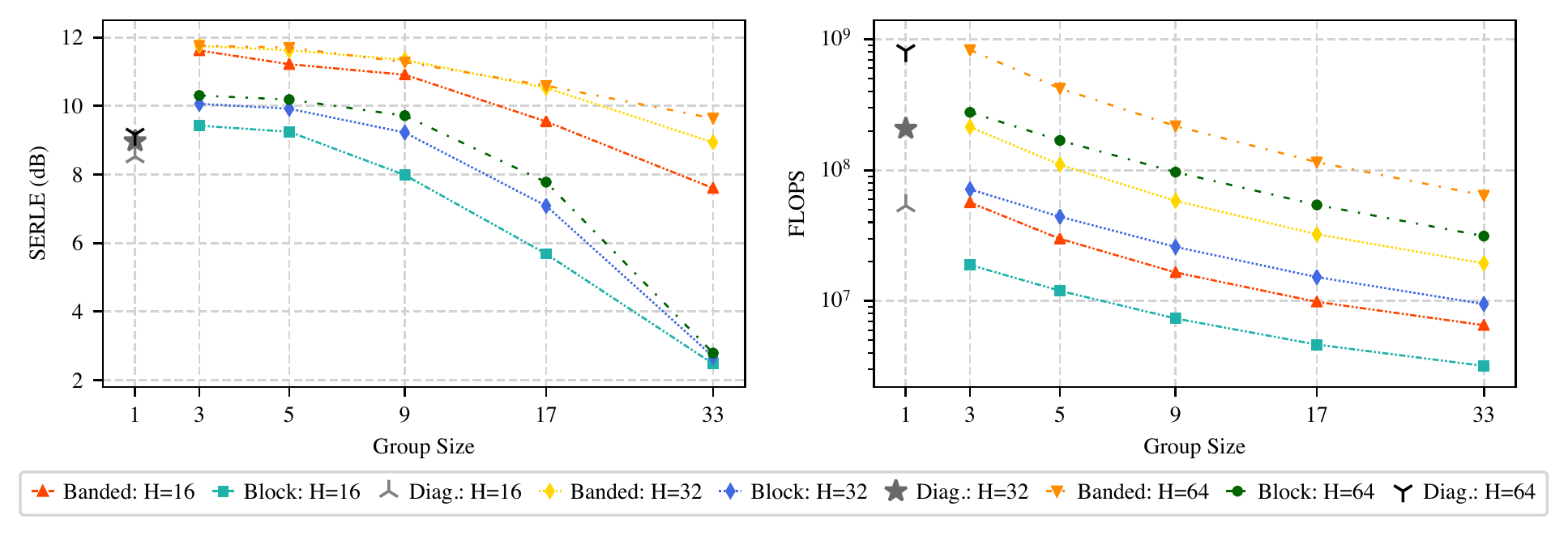}
	\vspace{-4.5mm}
	\caption{Higher-order frequency dependency comparison. Block and banded improve SERLE while reducing FLOPS.}
	\label{fig:serle_flops_compare}
	\vspace{-4.5mm}
\end{figure*}

\subsection{Training details}
For AEC, we use a $4096$ pt. window and a $2048$ pt. hop. We train Meta-AF models using \eqref{eq:metaloss} and the training scheme from~\cite{casebeer2022metaaf} with Adam~($\lambda=10^{-4}$). We train the DNN-SE using a $512$ pt. window, $256$ pt. hop, Adam~($\lambda=6 \cdot 10^{-4}$), with AEC-processed inputs and clean speech as the target using mean-squared error on the magnitude STFT. We set $L=20$ for Meta-AF training and $L=150$ for DNN-SE training. We use gradient clipping, $\frac{\lambda}{2}$ if val. performance does not improve for $5$ epochs, and stop training after $16$ with no improvement. 
\section{Results}


\subsection{Higher-order dependencies comparison}
We show the effect of different higher-order dependencies on performance~(left) and complexity~(right) in \fref{fig:serle_flops_compare}. We compare diagonal, block, and banded dependencies across GRU state sizes of $H=\{16, 32, 64\}$. Block and banded optimizers outperform their diagonal counterparts and reduce complexity. For diagonal optimizers, scaling $H$ has little impact on performance and increases complexity. However, for higher-order optimizers, scaling $H$ improves performance. For the higher-order optimizers, larger groups force more updates to be processed by a single GRU. For small groups this improves performance and reduces complexity. Intuitively, neighboring frequencies are related and grouping them allows the optimizer to exploit such relationships. All frequencies within a group share the same hidden state, which reduces the number of states, which reduces complexity. Overall, banded has the best SERLE but block is more efficient.

\subsection{Effect on downstream performance}
\begin{figure}[!t]
	\centering
	\includegraphics[trim=.25cm .25cm .25cm .25cm, clip, width=1\linewidth]{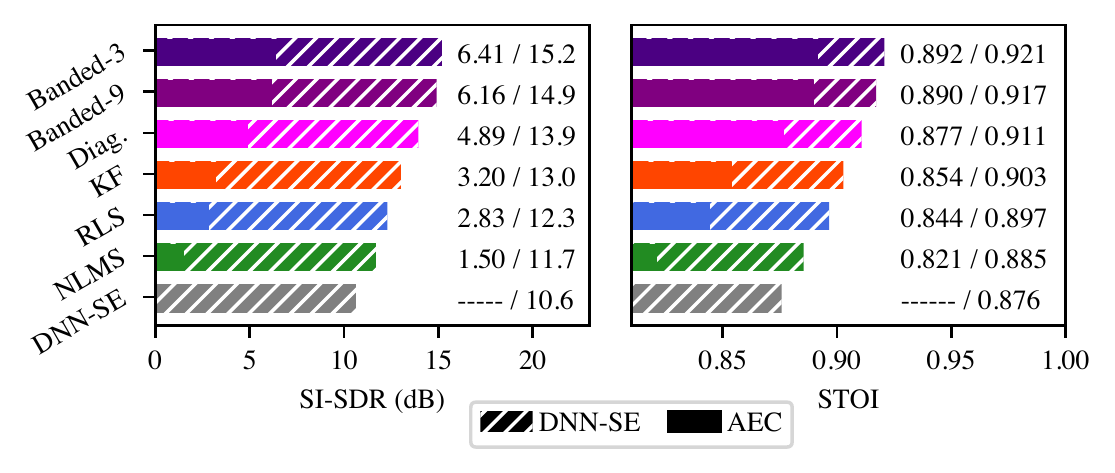}
	\vspace{-8.75mm}
	\caption{Effect of AEC on DNN-SE performance. Block Meta-AEC performs best before and after the DNN-SE.}
	\label{fig:res_compare}
	\vspace{-4mm}
\end{figure}

We show the effect of AEC on speech enhancement performance in~\fref{fig:res_compare}. In solid colors, we show performance of the AEC and in striped colors we show the performance of AEC along with a DNN-SE. We show three Meta-AEC models all with $H=32$: diagonal, banded with group $9$ and banded with group $3$. Banded-$3$ performs best and beats KF by $3.21\,$dB SI-SDR and $.038$ STOI. The less complex Banded-$9$ performs similarly and both beat diagonal Meta-AEC by $>1\,$dB SI-SDR and $>.01$ STOI. The trend holds when paired with a DNN-SE. Banded-$9$ surpasses KF by $2.2\,$dB SI-SDR and $.018$ STOI and diagonal by $1.3\,$dB SI-SDR and $.01$ STOI. This demonstrates that modeling higher-order dependencies translates to better downstream performance and highlights that AF advances can improve overall system performance. The raw mixture scores $-1.15\,$dB SI-SDR and $0.78\,$STOI. Oracle AEC and DNN-SE score $32.27\,$dB SI-SDR and $0.97\,$STOI. 

All AECs run in real-time on a single CPU core with RTFs of: $0.12$ for KF, $0.15$ for Diag., $0.18$ for Banded-$3$, and $0.13$ for Banded-$9$. Banded-$9$ is as fast as KF, and outperforms Diag. Meta-AEC models have $14$K complex parameters.

\section{Conclusion}
In this work, we propose a method for meta-learning adaptive filter update rules with higher order frequency dependencies. We evaluated a family of frequency dependency structures on a challenging acoustic echo cancellation task and found that our approach yields high performing and efficient update rules that run in real-time. We compared to a variety of competitive conventional and meta-learned AF baselines and show that our approach yields multi-dB improvements while being faster and less complex. Finally, we verify that our advances hold with and without a downstream speech enhancer.

\bibliographystyle{IEEEbib-abbrev}
\bibliography{refs}

\end{document}